\documentclass[a4paper]{article}
\usepackage{amssymb}
\usepackage{times,cite,bm}
\usepackage{diagram}

\newcommand{\be}{\begin{eqnarray}}
\newcommand{\ee}{\end{eqnarray}}
\newcommand{\bez}{\begin{eqnarray*}}
\newcommand{\eez}{\end{eqnarray*}}
\newcommand{\pa}{\partial}
\newcommand{\la}{\lambda}

\newcommand{\C}{\mathbb{C}}
\newcommand{\A}{\mathbb{A}}

\newcommand{\tr}{\mathrm{tr}}
\newcommand{\id}{\mathrm{id}}
\newcommand{\tpa}{\tilde{\pa}}
\newcommand{\Cdot}{\boldsymbol{\cdot}}
\newcommand{\hchi}{\hat{\chi}}
\begin{document}

\title{\bf Burgers and KP hierarchies: \\
           A functional representation approach
 \thanks{\copyright 2006 by A. Dimakis and F. M\"uller-Hoissen}
   }

\author{Aristophanes Dimakis \\
 Department of Financial and Management Engineering, \\
 University of the Aegean, 31 Fostini Str., GR-82100 Chios, Greece 
   \\ dimakis@aegean.gr
          \and
 Folkert M\"uller-Hoissen \\ Max-Planck-Institute for Dynamics 
 and Self-Organization \\
 Bunsenstrasse 10, D-37073 G\"ottingen, Germany 
   \\ folkert.mueller-hoissen@ds.mpg.de 
   }

\date{}

\renewcommand{\theequation} {\arabic{section}.\arabic{equation}}

\newtheorem{theorem}{Theorem}[section]
\newtheorem{lemma}{Lemma}[section]
\newtheorem{proposition}{Proposition}[section]
\newtheorem{definition}{Definition}[section]
\newtheorem{corollary}{Corollary}[section]

\maketitle

\begin{abstract}
From a `discrete' functional zero curvature equation, 
functional representations of (matrix) Burgers and potential KP 
(pKP) hierarchies (and others), as well as corresponding B\"acklund 
transformations, can be obtained in a surprisingly simple way. 
With their help we show that any solution of a 
Burgers hierarchy is also a solution of the pKP hierarchy. Moreover, 
the pKP hierarchy can be expressed in the form of an inhomogeneous 
Burgers hierarchy. In particular, this leads to an extension  
of the Cole-Hopf transformation to the pKP hierarchy. 
Furthermore, these hierarchies are solved by the 
solutions of certain functional Riccati equations. 
\end{abstract}

%\noindent
%{\bf Keywords:} Burgers hierarchy, Cole-Hopf transformation, KP hierarchy, functional Riccati equation

\section{Introduction}
\label{section:Intro}
\setcounter{equation}{0}
It has been noted in \cite{MSS91} (page~119) that any solution of the first two
equations of the Burgers hierarchy 
\cite{Choo+Choo77,LRB83,Brus+Ragn85,Pick94,Foka+Liu94,Tass96,Kupe00,Kupe05,DMH06nahier} 
is also a solution of the potential KP (pKP) equation. 
The generalization to the case where the dependent variables take their values 
in a matrix (or more generally an associative, and typically noncommutative) algebra 
$\A$ appeared in \cite{GMA94}. By use of functional representations (i.e., generating 
equations, depending on auxiliary indeterminates) of the corresponding 
hierarchies, it can easily be shown that indeed any solution of the (`noncommutative') 
Burgers hierarchy also solves the (`noncommutative') pKP hierarchy (see 
section~\ref{section:KP}). Moreover, it turns out that the pKP hierarchy 
can be expressed as an `inhomogeneous Burgers hierarchy'. This means that 
there is a functional form of the pKP hierarchy involving a matrix function 
as an inhomogeneous term. Setting the latter to zero, reduces it 
to a functional form of the Burgers hierarchy. 

Our starting point for the generation of functional representations of integrable 
hierarchies is a functional zero curvature (Zakharov-Shabat) equation, which 
we recall in section~\ref{section:funczc} (see also \cite{DMH06func,DMH06DNLS}). 
Section~\ref{section:Burgers} then treats the simplest non-trivial example, 
which is a Burgers hierarchy (with dependent variable in $\A$). Another version 
of the Burgers hierarchy is dealt with in appendix~A. 
Section~\ref{section:KP} addresses the case of the pKP hierarchy and its 
relations with Burgers hierarchies. In particular, we obtain an extension 
of the Cole-Hopf transformation from the Burgers to the pKP hierarchy, 
generalizing a result in \cite{GMA94}. 
Section~\ref{section:Riccati} shows that there are functional Riccati 
equations which imply the pKP, respectively Burgers hierarchy. Since 
such Riccati equations can be solved explicitly, 
this offers a quick way to exact solutions. Imposing a 
`rank one condition' (cf. \cite{Gekh+Kasm06} and references therein), 
these solutions of matrix hierarchies lead to solutions of the 
corresponding scalar hierarchies.

\section{The functional zero curvature condition}
\label{section:funczc}
\setcounter{equation}{0}
The integrability conditions of a linear system 
\be
    \pa_{t_n} \psi = B_n \psi \, , \qquad n=1,2,\ldots \, , 
\ee
with independent variables $\mathbf{t} := (t_1,t_2,t_3,\ldots)$, 
are the Zakharov-Shabat (zero curvature) conditions
\be
    \pa_{t_n} B_m - \pa_{t_m} B_n = [B_n,B_m] \, . \label{ZS}
\ee
We learned \cite{DMH06func,DMH06DNLS} that for several important hierarchies 
it is more convenient to use instead of the partial derivatives 
$\pa_{t_n}$ the operators
\be
       \hchi_n := p_n(-\tpa) \, , \qquad
       \tpa := (\pa_{t_1}, \pa_{t_2}/2, \pa_{t_3}/3, \ldots) \, ,
\ee
where $p_n$ are the elementary Schur polynomials, an insight which can be traced 
back to \cite{Sato+Sato82} (see also \cite{DJM82}).\footnote{In particular, 
we have $\hchi_0 = \id$, 
$\hchi_1 = - \pa_{t_1}$, $\hchi_2 = -{1\over2}\pa_{t_2} + {1\over2}\pa_{t_1}^2$, 
$\hchi_3 = -{1\over3}\pa_{t_3} + {1\over2}\pa_{t_2}\pa_{t_1} - {1\over6}\pa_{t_1}^3$, 
$\hchi_4 = -{1\over4}\pa_{t_4} + {1\over3}\pa_{t_3}\pa_{t_1} +{1\over8}\pa_{t_2}^2
    -{1\over4}\pa_{t_2}\pa_{t_1}^2 + {1\over24}\pa_{t_1}^4$.}
An equivalent form of the above linear system is then 
\be
   \psi_{-[\la]} = \mathcal{E}(\la) \, \psi \, ,  \label{funct_lin_sys}
\ee
where $\la$ is an indeterminate and $\mathcal{E}(\la) = \sum_{n \geq 0} \la^n \, \mathcal{E}_n$ a 
formal power series in $\la$.\footnote{The coefficients $\mathcal{E}_n$ can be expressed 
in terms of the $B_n$ and vice versa. For example, $B_1 = -\mathcal{E}_1$, 
$B_2 =-2 \mathcal{E}_2 - \mathcal{E}_{1,t_1} + \mathcal{E}_1^2$, 
$B_3 = -3\mathcal{E}_3 - 3\mathcal{E}_{2,t_1} - \mathcal{E}_{1,t_1t_1} + 2 \mathcal{E}_{1,t_1}\mathcal{E}_1 
 + \mathcal{E}_1 \mathcal{E}_{1,t_1} + 3 \mathcal{E}_2 \mathcal{E}_1 - \mathcal{E}_1^3$.}
Here we use the notation $[\la] := (\la,\la^2/2,\la^3/3,\ldots)$ and 
\be
    f_{-[\la]}(\mathbf{t}) := f(\mathbf{t}-[\la]) 
       = \sum_{n=0}^\infty \la^n \, \hchi_n(f) 
\ee
(as a formal power series in $\la$), for any object $f$ dependent on 
$\mathbf{t}$. This is sometimes refered to as a \emph{Miwa shift}.\footnote{We 
also use `positive' Miwa shifts, $f_{[\la]}(\mathbf{t}) := f(\mathbf{t}+[\la]) 
= \sum_{n=0}^\infty \la^n \, \chi_n(f)$ with $\chi_n := p_n(\tpa)$.} 
The integrability conditions now read
\be
    \mathcal{E}(\la)_{-[\mu]} \, \mathcal{E}(\mu)
  = \mathcal{E}(\mu)_{-[\la]} \, \mathcal{E}(\la) \; .  \label{EE}
\ee
with indeterminates $\la, \mu$. 
Regarding $\mathcal{E}(\la)$ as a parallel transport operator, 
(\ref{EE}) attains the interpretation of a `discrete' zero curvature 
condition, as depicted in the following (commutative) diagram.\footnote{Here 
`discrete' is used in the sense of \cite{DJM82}. See also \cite{ABS03,Bobe+Suri02} 
for an approach towards integrable equations via discrete zero curvature equations.}

\vskip-.8cm
\Large
\bez
\begin{diagram}
\node{\bullet} \arrow{e,t}{\mathcal{E}(\la)}
               \arrow{s,l}{\mathcal{E}(\mu)} 
\node{\bullet} \arrow{s,r}{\mathcal{E}(\mu)_{-[\la]}}  \\
\node{\bullet} \arrow{e,b}{\mathcal{E}(\la)_{-[\mu]}}
\node{\bullet} 
\end{diagram}
\eez
\normalsize
\vskip.2cm

Introducing a `discrete' gauge potential (cf. \cite{DMHS93a,DMHS93b}) via
\be
    \mathcal{E}(\la) = I - \la \, \mathcal{A}(\la) \, ,
\ee
where $I$ is the unit element of the (typically matrix) algebra 
from which the coefficients of the formal power series $\mathcal{A}(\la)$ 
are taken\footnote{More generally, the coefficients of the 
formal power series $\mathcal{E}(\la)$ and $\mathcal{A}(\la)$ may be elements of 
any unital associative algebra $\A$, the elements of which are differentiable 
with respect to the set of coordinates $\mathbf{t}$ (which requires a 
Banach space structure on $\A$). $\psi$ is then an element of a left $\A$-module.}, 
(\ref{EE}) can be written as
\be
    \Upsilon(\la,\mu) = \Upsilon(\mu,\la) \, ,  \quad
     \Upsilon(\la,\mu) 
  := \mu^{-1} (\mathcal{A}(\la) - \mathcal{A}(\la)_{-[\mu]}) 
     + \mathcal{A}(\la)_{-[\mu]} \, \mathcal{A}(\mu) \, .
      \label{zc_Ups}
\ee

Equation (\ref{EE}) exhibits the following gauge invariance,\footnote{(\ref{E-transf})
extends the above planar diagram to a `commutative cube' where $\mathcal{B}$ 
acts along the orthogonal bonds.}
\be
   \mathcal{E}(\la) \quad \mapsto \quad \mathcal{B}_{-[\la]} \, \mathcal{E}(\la) \, \mathcal{B}^{-1}
          = \mathcal{E}'(\la)     \label{E-transf}
\ee
with an invertible $\mathcal{B}$. This originates from the transformation
\be
    \psi' = \mathcal{B} \, \psi   \label{psi_transf}
\ee
of the linear system (\ref{funct_lin_sys}). In particular, B\"acklund 
(or Darboux) transformations arise in this way (see also \cite{CSY92}, for 
example). In terms of the gauge potential, (\ref{E-transf}) reads
\be
   \la^{-1} ( \mathcal{B} - \mathcal{B}_{-[\la]} ) 
 = \mathcal{A}'(\la) \, \mathcal{B} - \mathcal{B}_{-[\la]} \, \mathcal{A}(\la) \; .
         \label{A-transf}
\ee

In section~\ref{section:Burgers} a Burgers hierarchy results from the simplest 
non-trivial ansatz for $\mathcal{E}(\la)$ (see also appendix~A for another 
version of the matrix Burgers hierarchy). 
If the gauge potential is linear in the 
operator of partial differentiation with respect to a variable $x$, we obtain 
the pKP hierarchy, see section~\ref{section:KP}. 
There are more examples (see also \cite{DMH06func,DMH06DNLS}) and a generalization 
of (\ref{EE}) which covers multi-component hierarchies.

\section{The Burgers hierarchy, Cole-Hopf and B\"acklund transformation
         in functional form}
\label{section:Burgers}
\setcounter{equation}{0}
Choosing
\be
      \mathcal{E}(\la) = I - \la \, \phi \, ,   \label{E-Burgers}
\ee
so that $\mathcal{A}(\la) = \phi$ is independent of $\la$, then (\ref{EE}) 
can be expressed as
\be
   \omega(\la) = \omega(\mu) \, , \qquad 
   \omega(\la) := \la^{-1} (\phi-\phi_{-[\la]}) + \phi_{-[\la]} \phi \; .
        \label{funct_Burgers1}
\ee
Since $\lim_{\la \to 0} \omega(\la) = \phi_x + \phi^2$, where $x:=t_1$, 
this turns out to be equivalent to
\be
    \Omega(\la) 
 := \omega(\la) - \phi_x - \phi^2 
  \equiv ( \phi - \phi_{-[\la]} ) ( \la^{-1} - \phi ) - \phi_x
  = 0  \, ,   \label{funct_Burgers2}
\ee
which is a functional representation of a (`noncommutative') Burgers 
hierarchy. The first hierarchy equation is the Burgers equation
$\phi_y = \phi_{xx} + 2 \, \phi_x \phi$, where $y:=t_2$.\footnote{From 
(\ref{E-Burgers}) we obtain $B_1=\phi$, $B_2=\phi_{t_1} + \phi^2$, 
$B_3=\phi_{t_1t_1} + 2 \phi_{t_1} \phi + \phi \phi_{t_1} + \phi^3$, etc. 
The Zakharov-Shabat equations (\ref{ZS}) then also produce the 
Burgers hierarchy equations.}
\vskip.1cm

Since the curvature vanishes, we may expect that there is gauge in which the 
gauge potential $\mathcal{A}$ vanishes.
Hence, let us look for an invertible $f$ such that
\be
    f^{-1}_{-[\la]} \, \mathcal{E}(\la) \, f = I 
\ee
(i.e. $\mathcal{E}'(\la)=I$ and $\mathcal{B} = f^{-1}$ in (\ref{E-transf})), which is 
\be
    \la^{-1}(f - f_{-[\la]}) = \phi \, f  \; .    \label{funct_Hopf-Cole}
\ee

\begin{proposition}
\label{prop:HC}
(\ref{funct_Hopf-Cole}) is a functional representation of the Cole-Hopf transformation\footnote{This 
noncommutative version of the Cole-Hopf transformation (see 
\cite{Hopf50,Cole51,LRB83,WTC83,GMA94,Tass96,Jose+Vasu01,Arri+Hick02}, 
for example) for the Burgers equation appeared in \cite{LRB83,Arri+Hick02,Hama+Toda03b,Mart+Pash03}, for instance. }
\be
    \phi &=& f_x \, f^{-1}  \, ,    \label{Hopf-Cole}  \\
    \pa_{t_n} f &=& \pa_x^n f  \qquad n=2,3,\ldots \; . \label{heat_hier}
\ee
Any invertible $f$ which solves the linear `heat hierarchy' (\ref{heat_hier}) 
determines via (\ref{Hopf-Cole}) a solution of the Burgers 
hierarchy.\footnote{Conversely, if $\phi$ solves the Burgers hierarchy, 
choose $f$ such that $f_x = \phi \, f$. Then 
$0 = \Omega(\la) f = (\pa_x - \phi_{-[\la]})[ \la^{-1}(f - f_{-[\la]}) - f_x ]$ 
implies that $f$ solves the heat hierarchy if $\pa_x - \phi_{-[\la]}$ 
is invertible.}
\end{proposition}
{\it Proof:} A well-known identity for the elementary Schur polynomials $p_n$ leads to
\bez
     n \, \hchi_n
  = -\sum_{k=1}^n \pa_{t_k} \hchi_{n-k}
  = - \sum_{k=1}^{n-2} \pa_{t_k} \hchi_{n-k}
    - \pa_{t_n} + \pa_x \pa_{t_{n-1}}  \qquad n=2,3,\ldots \; .
\eez
With its help one proves by induction that (for an arbitrary integer $N>1$ 
the first $N$ of) the equations (\ref{heat_hier}) are equivalent to (the first 
$N$ of) the equations $\hchi_n(f) = 0$, $n=2,3,\ldots$. 
Together with (\ref{Hopf-Cole}), these equations are equivalent to  (\ref{funct_Hopf-Cole}). 
Furthermore, the integrability condition of (\ref{funct_Hopf-Cole}) is 
the Burgers hierarchy equation (\ref{funct_Burgers1}). 
\vskip.1cm

\noindent
{\bf Remark.} 
Special solutions of the heat hierarchy (\ref{heat_hier}) are given 
by arbitrary linear combinations of the Schur polynomials $p_n(\mathbf{t})$, 
$n = 0,1,2,\ldots$, with constant coefficients in $\A$. In particular, 
with constant $P \in \A$, the following is a (formal) solution,
\be
   e^{\xi(P)} = \sum_{n \geq 0} p_n(\mathbf{t}) P^n  \quad 
   \mbox{where} \quad \xi(P) := \sum_{m \geq 1} t_m P^m  \; . 
\ee 
\vskip.1cm

The transformation equation (\ref{A-transf}) now reads  
\be
    \la^{-1}(\mathcal{B} - \mathcal{B}_{-[\la]}) 
  = \phi' \, \mathcal{B} - \mathcal{B}_{-[\la]} \, \phi \; . \label{Burgers_preBT}
\ee
Taking $\la \to 0$, this implies
\be
   \phi' = \mathcal{B} \, \phi \, \mathcal{B}^{-1} 
          + \mathcal{B}_x \, \mathcal{B}^{-1} \; .  \label{Burgers_BTgauge}
\ee
Using the last equation to eliminate $\phi'$ from (\ref{Burgers_preBT}), yields
\be
   (\mathcal{B} - \mathcal{B}_{-[\la]})(\la^{-1} - \phi)  
 = \mathcal{B}_x  \; . \label{Burgers_BTcond}
\ee
Together with (\ref{Burgers_BTgauge}), this is equivalent to (\ref{Burgers_preBT}). 
Any invertible $\mathcal{B}$ which satisfies (\ref{Burgers_BTcond}), 
generates via (\ref{Burgers_BTgauge}) a new solution $\phi'$ from a given one 
$\phi$ of the Burgers hierarchy. 
Since (\ref{Burgers_BTcond}) is linear in $\mathcal{B}$, linear combinations 
of solutions (with constant left coefficients) are again solutions 
of (\ref{Burgers_BTcond}). 
Comparison with (\ref{funct_Burgers2}) shows that $\mathcal{B} = \phi$ is a 
particular solution. 
Obviously any constant element $\alpha$ also satisfies (\ref{Burgers_BTcond}). 
Hence $\mathcal{B} = \alpha + \beta \, \phi$, with arbitrary constant elements 
$\alpha, \beta$, satisfies these conditions and (\ref{Burgers_BTgauge}) 
takes the form
\be
  \phi' = (\alpha + \beta \, \phi) \, \phi \, (\alpha + \beta \, \phi)^{-1} 
          + \beta \, \phi_x \, (\alpha + \beta \, \phi)^{-1} \, ,  \label{Burgers_BT}
\ee
assuming the inverse to exist. 
This covers elementary B\"acklund (or Darboux) transformations obtained 
in \cite{Foka79,LRB83,WTC83,Kupe05} and \cite{Matv+Sall91}, p.73.

\section{The potential KP hierarchy in functional form, and relations with the Burgers hierarchy}
\label{section:KP}
\setcounter{equation}{0}
Choosing\footnote{Starting instead with $\mathcal{E}(\la) = I - \la v(\la) \pa$, 
leads in the same way to the modified KP hierarchy \cite{DMH06func}. 
The two choices of $\mathcal{E}(\la)$ are related by a gauge transformation 
(Miura transformation).}
\be
    \mathcal{E}(\la) = I - \la \, (w(\la) + \pa) \, ,
\ee
i.e. $\mathcal{A}(\la) = w(\la) + \pa$, where $\pa = \pa_x$, 
(\ref{zc_Ups}) leads to the two equations
\be
  &&  \la^{-1} (w(\mu) - w(\mu)_{-[\la]}) 
     + w(\mu)_{-[\la]} \, w(\la) + w(\la)_x  \\
  &=& \mu^{-1} (w(\la) - w(\la)_{-[\mu]}) 
     + w(\la)_{-[\mu]} \, w(\mu) + w(\mu)_x
\ee
and
\be
    w(\la) - w(\la)_{-[\mu]}  = w(\mu) - w(\mu)_{-[\la]}  \; .
\ee
The latter is solved by
\be
    w(\la) = \phi -\phi_{-[\la]} \, , 
\ee
and the first equation then takes the form
\be
    \omega(\la)_{-[\mu]} - \omega(\mu)_{-[\la]}
  = \omega(\la) - \omega(\mu) -(\phi_x + \phi^2)_{-[\la]}
     + (\phi_x+\phi^2)_{-[\mu]} \, ,  \label{funct_KP}
\ee
using the definition in (\ref{funct_Burgers1}). 
Summing this expression three times with cyclically permuted indeterminates 
$\la_1,\la_2,\la_3$, results in the Bogdanov-Konopelchenko (BK) functional 
equation \cite{Bogd+Kono98,Bogd99},  
\be
    \sum_{i,j,k=1}^3 \epsilon_{ijk} \, \omega(\la_i)_{-[\la_j]} = 0  \label{KP-BK}
\ee
(where $\epsilon_{ijk}$ is totally antisymmetric with $\epsilon_{123}=1$). 
This determines the pKP hierarchy and is equivalent to (\ref{funct_KP}). 
Expanding (\ref{funct_KP}) in  $\la, \mu$, yields $\pa_x \phi = \pa_{t_1} \phi$ 
and 
\be
    \hchi_m \hchi_{n+1}(\phi) - \hchi_n \hchi_{m+1}(\phi) 
  = \hchi_m (\hchi_n(\phi) \, \phi) - \hchi_n (\hchi_m(\phi) \, \phi) 
    \quad  m,n=1,2, \ldots \; . 
    \label{KP_hchi}
\ee
An equivalent expression of the pKP hierarchy (in the scalar case) 
appeared already in \cite{DNS89} (see also \cite{DMH06nahier,DMH06func}). 
For $m=1, n=2$, this yields the pKP equation
\be
   ( 4 \, \phi_t - \phi_{xxx} - 6 \, \phi_x{}^2 )_x - 3 \, \phi_{yy} 
   + 6 [ \phi_x , \phi_y ] = 0 \, ,   \label{pKPequation}
\ee
where $x=t_1, y=t_2, t=t_3$.
Comparing (\ref{funct_Burgers1}) with (\ref{funct_KP}) shows that 
any solution of the Burgers hierarchy, considered in section~\ref{section:Burgers},  
also solves the pKP hierarchy.
\vskip.1cm

\noindent
{\bf Remark.} There is a (Sato-Wilson) pseudo-differential operator
$W = I + \sum_{n>0} w_n \pa^{-n}$ such that 
$\mathcal{B} = W^{-1}$ in (\ref{E-transf}) transforms 
$\mathcal{E}(\la)$ to $\mathcal{E}'(\la) = I - \la \pa$. 
It is determined (up to multiplication by a constant operator 
$I + \sum_{n>0} c_n \pa^{-n}$) by 
$w_1 - w_{1,-[\la]} = \phi_{-[\la]} - \phi$ and 
$w_{n+1}-w_{n+1,-[\la]} = \la^{-1}(w_n - w_{n,-[\la]})-w_{n,x} 
 - (\phi-\phi_{-[\la]}) \, w_n$.

\subsection{The pKP hierarchy as an inhomogeneous Burgers hierarchy}
We observe that (\ref{funct_KP}) can also be written as
\be
     \Omega(\mu) - \Omega(\mu)_{-[\la]} 
   = \Omega(\la) - \Omega(\la)_{-[\mu]}
\ee
where $\Omega(\la)$ is the expression defined in (\ref{funct_Burgers2}) 
in terms of $\phi$. As a consequence, the pKP hierarchy takes the form
\be
    \Omega(\la) = \theta - \theta_{-[\la]} \; .  \label{KP:Omega->theta}
\ee
with some $\theta$. 
If the right hand side vanishes, i.e. if $\theta$ is constant, this is 
precisely the functional representation (\ref{funct_Burgers2}) of the 
Burgers hierarchy considered in section~\ref{section:Burgers}. 
(\ref{KP:Omega->theta}) is equivalent to
\be
   \hchi_{n+1}(\phi) - \hchi_n(\phi) \, \phi = \hchi_n(\theta) 
   \qquad n=1,2, \ldots \; .
\ee
The first two equations are 
\be
 &&  \phi_y = \phi_{xx} + 2 \, \phi_x \, \phi + 2 \, \theta_x \, , 
         \label{1stKP-Burgers} \\
 &&  \phi_t = \phi_{xxx} + 3 \phi_{xx} \phi + 3 \phi_x{}^2  
    + 3 \, \phi_x \, \phi^2 + 3 \, \theta_x \, \phi 
    + \frac{3}{2} (\theta_y + \theta_{xx} ) \, ,   \label{2ndKP-Burgers}
\ee
where we used the first to replace $\phi_y$ in the second equation. 
For constant $\theta$, these are the first two equations of the Burgers 
hierarchy. Eliminating $\theta$ from (\ref{1stKP-Burgers}) and 
(\ref{2ndKP-Burgers}), we recover the pKP equation (\ref{pKPequation}). 
\vskip.1cm

Application of a Miwa shift to (\ref{funct_KP}) leads to
\be
    \tilde{\omega}(\la)_{[\mu]} - \tilde{\omega}(\mu)_{[\la]}
  = \tilde{\omega}(\la) - \tilde{\omega}(\mu) - (\phi_x+\phi^2)_{[\la]}
     + (\phi_x+\phi^2)_{[\mu]} \, , 
\ee
where
\be
   \tilde{\omega}(\la) := \omega(\la)_{[\la]}
  = \la^{-1} (\phi_{[\la]} - \phi) + \phi \, \phi_{[\la]} \; .
\ee
Since this can be written as
\be
   \tilde{\Omega}(\la)_{[\mu]} - \tilde{\Omega}(\la)
 = \tilde{\Omega}(\mu)_{[\la]} - \tilde{\Omega}(\mu) 
\ee
with
\be
    \tilde{\Omega}(\la) := \tilde{\omega}(\la) - \phi_x - \phi^2
  = (\la^{-1} + \phi)(\phi_{[\la]} - \phi) - \phi_x \, , 
\ee 
the pKP hierarchy can also be expressed as
\be
   \tilde{\Omega}(\la) = \tilde{\theta}_{[\la]} - \tilde{\theta}
\ee
with some $\tilde{\theta}$.\footnote{$\tilde{\theta}$ and $\theta$ are 
related by $\tilde{\theta} - \theta = \phi_x + \phi^2$.} 
If $\tilde{\theta}$ is constant, so that the 
right hand side vanishes, the last equation reduces to the 
`opposite' Burgers hierarchy (see also appendix~A), 
\be
   (\la^{-1} + \phi)(\phi_{[\la]} - \phi) = \phi_x \, , 
          \label{oppBurgers}
\ee 
which starts with $\phi_y = - \phi_{xx} - 2 \, \phi \, \phi_x$. 
In particular, we have the following result. 
\vskip.1cm

\begin{proposition}
Any solution of any of the two Burgers hierarchies also solves the 
pKP hierarchy.
\end{proposition}

\subsection{A Cole-Hopf transformation for the matrix pKP hierarchy}

\begin{theorem}
\label{theorem:pKP-HC}
Let $(\A,\Cdot)$ be the algebra of $M \times N$ matrices of functions 
of $\mathbf{t}$ with the product 
\be
        A \Cdot B = A Q B \, ,  \label{Qproduct}
\ee
where the ordinary matrix product is used on the right hand side, and 
$Q$ is a constant $N \times M$ matrix. 
Let $X$ be an invertible $N \times N$ matrix and $Y \in \A$, such that 
$X,Y$ solve the linear heat hierarchy (\ref{heat_hier}) and 
satisfy
\be
    X_x = R \, X + Q \, Y \, ,   \label{X_x}
\ee
with a constant $N \times N$ matrix $R$. The pKP hierarchy in $(\A,\Cdot)$ 
is then solved by
\be 
      \phi := Y X^{-1} \; . \label{pKP-HC}
\ee
\end{theorem}
{\it Proof:}  By use of (\ref{pKP-HC}), the expression $\Omega(\la)$ 
defined in (\ref{funct_Burgers2}) (where because of (\ref{Qproduct}) 
a factor $Q$ enters the nonlinear term) can be written as
\bez
    \Omega(\la) &=& ( \phi -\phi_{-[\la]} )( X_x  - Q Y ) \, X^{-1}
    + ( \la^{-1} ( Y-Y_{-[\la]} ) - Y_x ) \, X^{-1} \\
   && - \phi_{-[\la]} ( \la^{-1} ( X - X_{-[\la]} ) - X_x ) \, X^{-1} \; .
\eez
If $X,Y$ solve the heat hierarchy, then 
$\hchi_{n}(X) = 0 = \hchi_{n}(Y)$, $n=2,3,\ldots$, and thus
\bez
    \la^{-1} (X-X_{-[\la]}) = X_x \, , \qquad 
    \la^{-1} (Y-Y_{-[\la]}) = Y_x \; .  
\eez
Using these equations, the above expression for $\Omega(\la)$ reduces to
\bez
   \Omega(\la) = ( \phi - \phi_{-[\la]})(X_x - Q Y) X^{-1} \; .
\eez
If $(X_x - Q Y) X^{-1}$ is constant, say $R$, which means that 
(\ref{X_x}) holds, this takes the form (\ref{KP:Omega->theta}) of the 
pKP hierarchy with $\theta = \phi \, R$.\footnote{Note also that 
$\tilde{\theta} = \theta + \phi_x + \phi Q \phi = Y_x X^{-1}$
by use of (\ref{X_x}).} 
Thus $\phi$ solves the pKP hierarchy.
\hfill $\blacksquare$
\vskip.1cm

If $R=0$, and with $M=N$ and $Q=I_N$, (\ref{X_x}) and (\ref{pKP-HC}) 
reduce to $\phi = X_x X^{-1}$, and we recover the Cole-Hopf transformation 
for the Burgers hierarchy. Note that the conditions imposed on $X$ already 
imply $Q(\la^{-1}(Y-Y_{-[\la]}) - Y_x)=0$ and thus $Y$ automatically 
satisfies the heat hierarchy if $Q$ has maximal rank. Furthermore, if we 
consider $Q \phi$ instead of $\phi$, the assumption on $Y$ can be dropped. 
\vskip.1cm

\begin{corollary}
Let $X$ solve the heat hierarchy and (\ref{X_x}) with some $Y$. Then $Q \phi$ 
with $\phi$ given by (\ref{pKP-HC}) solves the $N \times N$ matrix pKP hierarchy 
with the usual matrix product.
\end{corollary}

 For the case where $\mathrm{rank}(Q)=1$ (cf. \cite{Gekh+Kasm06} and 
references therein), a similar result appeared already in \cite{GMA94}. 
Then $\tr(Q A \Cdot B) = \tr(Q A) \, \tr(Q B)$, so that, by use of 
(\ref{X_x}),
\be
    \varphi := \tr(Q \phi) = -\tr(R) + (\log\tau)_x \quad
    \mbox{with} \quad
    \tau := \det(X)     \label{varphi,tau}
\ee
solves the \emph{scalar} pKP hierarchy.

\subsection{B\"acklund and Darboux transformations}
Inserting the ansatz $\mathcal{B} = b(\mathbf{t}) - \pa$ in (\ref{A-transf}) 
leads to the two equations
\be
    b - \phi' + \phi = ( b - \phi' + \phi )_{-[\la]} \label{KP-BD_1}
\ee
and
\be
    \la^{-1}(b-b_{-[\la]}) - b_x = (\phi' - \phi'_{-[\la]}) b - b_{-[\la]}
    (\phi-\phi_{-[\la]})+(\phi-\phi_{-[\la]})_x \; . \label{KP-BD_2}
\ee
The solution of (\ref{KP-BD_1}) is
\be
    b = \phi' - \phi    \label{KP:b->phi}
\ee
(absorbing an additive constant into $\phi'$).  
(\ref{KP-BD_2}) can then be written as 
\be
   \Omega(\la) - \Omega'(\la) = \Gamma(\phi,\phi') - \Gamma(\phi,\phi')_{-[\la]} \, ,
\ee
where $\Omega'(\la)$ is built with $\phi'$, and
\be
    \Gamma(\phi,\phi') := (\phi' - \phi) \, \phi - \phi_x \; .
\ee
This is an elementary B\"acklund transformation (BT) of the pKP 
hierarchy.\footnote{Extending the above ansatz for $\mathcal{B}$ 
to $n$th order in $\pa$ leads to equations which determine $n$th order 
BTs. These are solved by an $n$-fold product of elementary BTs.}
Using (\ref{KP:Omega->theta}), we find
\be
   0 = \Gamma(\phi,\phi') + \theta' - \theta 
     = \phi' \phi + \theta' - \tilde{\theta}  \; .
\ee
Let $\mathcal{B}_{n,m}$ denote the BT taking a pKP 
solution $\phi_m$ to a new solution $\phi_n$. The permutability 
relation\footnote{Note that this is also a discrete zero curvature condition.}
$\mathcal{B}_{(3,1)} \mathcal{B}_{(1,0)} = \mathcal{B}_{(3,2)} \mathcal{B}_{(2,0)}$  
then results in\footnote{In the commutative scalar case, setting 
$\phi=\tau_x/\tau$ with a function $\tau$, yields 
$\tau_0 \tau_3 = \tau_1 \tau_{2,x} - \tau_{1,x} \tau_2$.} 
\be
      (\phi_2-\phi_1)_x 
   = \phi_3 (\phi_2 - \phi_1) + (\phi_2 - \phi_1) \phi_0   
     + \phi_1^2 - \phi_2^2 \; .
\ee
This determines algebraically a new solution $\phi_3$ in terms of a 
given solution $\phi_0$ and corresponding B\"acklund descendants $\phi_1,\phi_2$. 
\vskip.1cm

In the case under consideration, the linear system (\ref{funct_lin_sys}) 
takes the form
\be
    \la^{-1} (\psi-\psi_{-[\la]}) - \psi_x = (\phi-\phi_{-[\la]}) \psi 
\ee
(cf. \cite{Sato+Sato82} for an equivalent version in the scalar case). 
If $\psi$ is invertible, we obtain
\be
    \phi - \phi_{-[\la]} 
  = \la^{-1} (\psi - \psi_{-[\la]}) \psi^{-1} - \psi_x \psi^{-1} \; .
\ee
Eliminating $\phi'$ from (\ref{KP-BD_2}) with the help 
of (\ref{KP:b->phi}), and then $\phi-\phi_{-[\la]}$ by use of the last equation, 
turns it into
\be
      (b - \psi_x \psi^{-1})_x 
    + (b - \psi_x \psi^{-1})(b + \la^{-1} \psi_{-[\la]} \psi^{-1}) &&  \nonumber \\
    - (b_{-[\la]} + \la^{-1} \psi_{-[\la]} \psi^{-1})(b - \psi_x \psi^{-1}) &=& 0 \; .
\ee
This equation is obviously solved by
\be
    b = \psi_x \psi^{-1} \; .
\ee
Hence, if $\psi_1$ solves the linear system with a solution $\phi$ 
of the pKP hierarchy, then 
\be
      \phi' = \phi + \psi_{1,x} \psi_1^{-1} 
\ee
is a new solution of the pKP hierarchy.\footnote{Moreover,
$\psi' = \mathcal{B} \psi = \psi_x - \psi_{1,x} \psi_1^{-1} \psi$ 
satisfies the linear system with $\phi'$.} 
This is a Darboux transformation \cite{Matv79KP,Matv+Sall91,Oeve93,Liu+Mana99}.

\section{Functional Riccati equations associated with KP and Burgers hierarchies, and 
exact solutions}
\label{section:Riccati}
\setcounter{equation}{0}
Let us consider the BK functional equation (\ref{KP-BK}) 
in the algebra $(\A, \Cdot)$, where $\A$ is the set of $M \times N$ 
matrices of complex functions of $\mathbf{t}$, supplied with the product 
(\ref{Qproduct}). 
The simplest non-trivial equation, which results from this formula by 
expansion in powers of the indeterminates, is the matrix pKP equation 
\be
    (4 \phi_t -\phi_{xxx} - 6 \phi_x Q \phi_x)_x 
   = 3 \phi_{yy} - 6(\phi_x Q \phi_y - \phi_y Q \phi_x) \; .
\ee
As a consequence, $\phi Q$ satisfies the $M \times M$ matrix pKP hierarchy and 
$Q \phi$ the $N\times N$ matrix pKP hierarchy. Moreover, if $Q = V U^T$, with 
an $N \times m$ matrix $V$ and an $M \times m$ matrix $U$, then $U^T \phi V$ satisfies 
the $m \times m$ matrix pKP hierarchy. In particular, for $m=1$ this becomes
the scalar pKP hierarchy. In the latter case, $Q$ has rank one. 
\vskip.1cm

The crucial observation now is that the BK functional equation, and thus 
the pKP hierarchy, is satisfied if $\phi$ solves
\be
    \omega(\la) = S + L \phi - \phi_{-[\la]} R  \label{Riccati_1}
\ee
with constant matrices $S , L, R$ of dimensions $M \times N$, $M \times M$ 
and $N \times N$, respectively. 
This is a functional matrix Riccati equation for $\phi$, 
\be
    \la^{-1}(\phi - \phi_{-[\la]}) 
  = S + L \phi - \phi_{-[\la]} R - \phi_{-[\la]} Q \phi \; . \label{Riccati_2}
\ee
The integrability condition of this functional equation is 
satisfied\footnote{This also follows from our work in 
\cite{DMH06nahier} and is the reason for the choice of the right hand 
side of (\ref{Riccati_1}).}, since
\be
 & & (\phi_{-[\la]})_{-[\mu]}
  = [ (\la^{-1}-L)\phi_{-[\mu]} -S ] [ (\la^{-1} -  R)
        - Q \phi_{-[\mu]} ]^{-1}  \nonumber \\
 &=& [ (\la^{-1}-L)(\mu^{-1}-L)\phi -(\la^{-1} -\mu^{-1})S + L S + S R
        +S Q \phi ] \nonumber \\
 & & \times [ (\la^{-1}-R)(\mu^{-1}-R)-(\la^{-1}+\mu^{-1})Q\phi
        + (R Q + Q L)\phi +Q S ]^{-1} \quad
\ee
is symmetric in $\la, \mu$ and thus equals $(\phi_{-[\mu]})_{-[\la]}$. 
The Riccati equation implies
\be
   \Omega(\la) = (\phi - \phi_{-[\la]}) \, R \, , \qquad
  \tilde{\Omega}(\la) = L \, (\phi_{[\la]} - \phi) \; .
\ee
This shows that with $R=0$ (respectively $L=0$), any solution 
of (\ref{Riccati_2}) also solves the Burgers hierarchy (\ref{funct_Burgers2}) 
(respectively the opposite Burgers hierarchy (\ref{oppBurgers})) in $(\A, \Cdot)$. 
\vskip.1cm

It is well-known that matrix Riccati equations can be 
linearized \cite{Reid72,AFIJ03}.\footnote{This is achieved by regarding 
$\phi(\mathbf{t})$ as an element of the Grassmannian $G(N,N+M)$ of 
$N$-dimensional linear subspaces of 
$\C^{N+M}$ via $\kappa(\phi) = \mathrm{span}(I_N,\phi^T)^T$, 
since $\kappa^{-1} : G(N,N+M) \rightarrow \C^{M \times N}$ defines a chart
for the manifold $G(N,N+M)$.} 
In fact, (\ref{Riccati_2}) can be lifted to a linear equation 
on the space of $(N+M) \times N$ matrices:
\be
    \la^{-1} (Z - Z_{-[\la]}) = H Z    \label{Riccati_linsys}
\ee
with
\be
    Z = \left(\begin{array}{c} X \\ Y \end{array}\right) \, , \qquad
    H = \left(\begin{array}{cc} R & Q \\ S & L \end{array}\right) \; .
\ee
Hence
\be
    \la^{-1} (X-X_{-[\la]}) = R X + Q Y \, , \qquad
    \la^{-1} (Y-Y_{-[\la]}) = S X + L Y \; .    \label{XYeqs}
\ee
Assuming that $X$ is invertible and setting 
\be
      \phi = Y \, X^{-1} \, ,    \label{phi=YX^-1}
\ee
these equations imply
\be
        \phi_{-[\la]} 
     = Y_{-[\la]} X_{-[\la]}^{-1} 
     = [ \phi - \la ( S + L \phi)] [ I_N - \la (R + Q \phi) ]^{-1} \, ,
\ee
which is (\ref{Riccati_2}). 
Thus any solution $Z$ of the linear functional equation 
(\ref{Riccati_linsys}) with invertible $X$ determines via (\ref{phi=YX^-1}) a 
solution of the functional matrix Riccati equation (\ref{Riccati_2}), and thus 
a solution of the matrix pKP hierarchy we started with. 
\vskip.1cm

\noindent
{\bf Remark.} The first of equations (\ref{XYeqs}) is equivalent to 
(\ref{X_x}) and the heat hierarchy for $X$. Since the second of 
(\ref{XYeqs}) implies that also $Y$ has to solve the heat hierarchy, 
according to theorem~\ref{theorem:pKP-HC} the $\phi$ determined by 
(\ref{phi=YX^-1}) already solves the pKP hierarchy without use of 
the additional equation $Y_x = S X + L Y$ which results from 
the second of (\ref{XYeqs}). However, this equation helps to single 
out interesting classes of solutions, see below. In any case, the Riccati 
approach corresponds to a class of (generalized) Cole-Hopf transformations 
in the sense of theorem~\ref{theorem:pKP-HC}. 
Note also that $\tilde{\theta}= S + L \phi$. 
\vskip.1cm

The general solution of (\ref{Riccati_linsys}) is
\be
    Z = e^{\xi(H)} Z_0 \, , \qquad \xi(H) = \sum_{n \geq 1} H^n t_n, \qquad
    Z_0 = \left(\begin{array}{c} X_0 \\ Y_0 \end{array}\right) \, ,
\ee
with invertible $X_0$. As a consequence, $Z_{t_n} = H^n Z$. 
Writing
\be
  e^{\xi(H)} =: \left(\begin{array}{cc} \Xi_{11} & \Xi_{12} \\ \Xi_{21} & \Xi_{22} 
         \end{array}\right)  \, ,
\ee 
we have
\be
    \phi = (\Xi_{21} + \Xi_{22} \, \phi_0)(\Xi_{11} + \Xi_{12} \, \phi_0)^{-1} \, , 
             \label{KP_frac_trans}
\ee
where $\phi_0 = Y_0 X_0^{-1}$. 
This is a matrix fractional transformation with coefficients depending on $\mathbf{t}$. 
For any choice of the matrices $S,L,R,Q$, this $\phi$ is a solution of the  
pKP hierarchy in the matrix algebra with product (\ref{Qproduct}). 
The practical problem is to compute $e^{\xi(H)}$ explicitly. 
\vskip.1cm

\noindent
{\bf Remark.} With $Z = e^{\xi(H)} Z_0$ also $T Z$ satisfies (\ref{Riccati_linsys}) 
if $T$ is constant and commutes with $H$. 
In particular, $T=k I_{M+N} + H$ with any constant $k$ induces such a transformation. 
It results in the matrix fractional transformation (with constant coefficients)
$\phi' = (S + L' \phi)(R' + Q \phi)^{-1}$ with $L' := L + k I_M$, $R' := R + k I_N$. 
\vskip.1cm

\noindent
{\bf Example 1.} Let $S=0$ and 
\be
    Q = R K - K L 
\ee
with a constant $N \times M$ matrix $K$. Then we have
\be
    H^n = \left(\begin{array}{cc} R^n & R^n K -K L^n \\ 0 & L^n \end{array}\right) \, , 
    \quad
    \xi(H) = \left(\begin{array}{cc} \xi(R) & \xi(R) K - K \xi(L) \\ 
                                      0 & \xi(L) \end{array}\right) \, , 
\ee
and thus 
\be
    e^{\xi(H)} = \left(\begin{array}{cc} e^{\xi(R)} & e^{\xi(R)} K -K e^{\xi(L)} \\ 
                                         0 & e^{\xi(L)} \end{array}\right) \, ,
\ee
so that (\ref{KP_frac_trans}) becomes
\be
    \phi = e^{\xi(L)} \phi_0 (I_N + K \phi_0 
      - e^{-\xi(R)} K e^{\xi(L)}\phi_0)^{-1} e^{-\xi(R)} \; .
\ee
If $Q$ has rank one, then we obtain the following solution of the scalar 
pKP hierarchy,
\be
    \varphi 
 &=& \tr(Q \phi) 
  = \tr \log( I_N + K \phi_0 -e^{-\xi(R)} K e^{\xi(L)}\phi_0 )_x 
  = (\log \tau)_x \, , \qquad \\
    \tau &=& \det(I_N + K \phi_0 - e^{-\xi(R)} K e^{\xi(L)}\phi_0) \, , 
\ee
which includes well-known formulae for KP multi-solitons \cite{MZBIM77} 
and resonances (see e.g. \cite{Bion+Chak06,DMH06CJP} and references therein). 
\vskip.1cm

\noindent
{\bf Example 2.}
Let $M=N$ and
\be
    L = S \pi_{-} \, , \qquad R = \pi_{+} S \, , \qquad
    Q = \pi_{+} S \pi_{-} \, ,
\ee
with constant $N \times N$ matrices $S, \pi_{\pm}$ such that $\pi_{+} + \pi_{-} = I_N$. 
It is easy to see that
\be
    H^n = \left(\begin{array}{cc} \pi_{+} S^n & \pi_{+} S^n \pi_{-} \\
    S^n & S^n \pi_{-} \end{array}\right) \; .
\ee
As a consequence, we find
\be
    e^{\xi(H)} =\left(\begin{array}{cc} 
     \pi_{-} + \pi_{+} e^{\xi(S)} & \pi_{+}(e^{\xi(S)}-I_N)\pi_{-} \\
     e^{\xi(S)} - I_N & \pi_{+} + e^{\xi(S)} \pi_{-} \end{array}\right) \, ,
\ee
and (\ref{KP_frac_trans}) reads
\be
    \phi = (-A + e^{\xi(S)} B)(\pi_{-} A + \pi_{+} e^{\xi(S)} B)^{-1} \, ,
\ee
where $A := I_N - \pi_{+} \phi_0$, $B := I_N + \pi_{-} \phi_0$.
If $\mathrm{rank}(\pi_{+} S \pi_{-}) = 1$, then 
\be
    \varphi = \tr(Q \phi) = -\tr(\pi_{+}S) + (\log\tau)_x \, , \qquad
    \tau = \det(\pi_{-} A + \pi_{+} e^{\xi(S)} B) \; .
\ee
 For example, let $N = m+n$ and choose
\be
    \pi_{-} = \left(\begin{array}{cc} I_m & 0 \\ 0 & 0 \end{array}\right) \, , \qquad
    \pi_{+} = \left(\begin{array}{cc} 0 & 0 \\ 0 & I_n \end{array}\right) \; . 
\ee
Writing
\be
    \phi_0 = \left(\begin{array}{cc} (\phi_0)_{--} & (\phi_0)_{-+} \\ 
              (\phi_0)_{+-} & (\phi_0)_{++} \end{array}\right) \, , \qquad
    S = \left(\begin{array}{cc} S_{--} & S_{-+} \\ S_{+-} & S_{++} \end{array}\right) 
        \, , 
\ee
$\mathrm{rank}(Q)=1$ means $\mathrm{rank}(S_{+-})=1$  (see also \cite{Gekh+Kasm06}) 
and we find
\be
    \tau = \det((e^{\xi(S)})_{++} + (e^{\xi(S)})_{+-}(\phi_0)_{-+}) \, .
\ee
In particular, if $S$ is the shift operator $S e_i = e_{i+1}$, 
this determines $\tau$-functions which can be expressed in terms 
of Schur polynomials. This corresponds to a finite version of the 
Sato theory, see \cite{Gekh+Kasm06}. 
For example, if $m=n=2$ and 
\be
    (\phi_0)_{-+} = \left(\begin{array}{cc} a & b \\ c & d \end{array}\right) \, , 
\ee
we obtain
\be
 \tau &=& 1 + c x + a \Big( y +{x^2 \over 2} \Big) + d \Big( y - {x^2 \over 2} \Big) 
          + b \Big( t-{x^3 \over 3} \Big)  \nonumber \\
      & & + (a d - b c) \Big( -x t + y^2 + {x^4 \over 12} \Big) \; .
\ee

\renewcommand{\thesection} {\Alph{section}}
\renewcommand{\theequation} {\Alph{section}.\arabic{equation}}
\renewcommand{\thelemma} {\Alph{section}.\arabic{lemma}}
\renewcommand{\theproposition} {\Alph{section}.\arabic{proposition}}
\renewcommand{\thecorollary}{\Alph{section}.\arabic{corollary}}
\setcounter{section}{1}

\section*{Appendix A: Opposite Burgers hierarchy and beyond}
\setcounter{equation}{0}
We generalize the ansatz for $\mathcal{E}(\la)$ considered in section~\ref{section:Burgers} to
\be
      \mathcal{E}(\la) = I - \la \, \sum_{n \geq 0} \la^n \, \phi_n \; .
\ee
Then (\ref{zc_Ups}) takes the form
\be
    \hchi_{n+1}(\phi_m) - \hchi_{m+1}(\phi_n) 
  = \sum_{k=0}^n \hchi_k(\phi_m) \, \phi_{n-k}
     -\sum_{k=0}^m \hchi_k(\phi_n) \, \phi_{m-k} \, ,   \label{big_hier}
\ee
where $m,n = 0,1,2, \ldots$. This is an infinite system of coupled equations. 
As in section~\ref{section:Burgers}, we look for a gauge transformation 
such that $f^{-1}_{-[\la]} \, \mathcal{E}(\la) \, f = I$, which is 
\be
    \la^{-1} (f - f_{-[\la]}) = \sum_{n \geq 0} \la^n \, \phi_n \, f  \; .   
          \label{generalized_Hopf-Cole}
\ee
Expanding the left hand side in powers of $\la$, this becomes
a generalization of the Cole-Hopf transformation,
\be
    \phi_0 = f_x \, f^{-1} \, ,  \qquad
    \phi_n = - \hchi_{n+1}(f) \, f^{-1} \qquad n=1,2, \ldots \; .  \label{gHC}
\ee
By construction, this solves the zero curvature equation and thus 
the hierarchy (\ref{big_hier}). 
The gauge transformation (\ref{A-transf}) takes the form
\be
    \la^{-1} ( \mathcal{B} - \mathcal{B}_{-[\la]} )
  = \sum_{n=0}^\infty \la^n \, (\phi_n' \, \mathcal{B} 
      - \mathcal{B}_{-[\la]} \, \phi_n )  \, ,
\ee
and thus
\be
   \phi_0' &=& \mathcal{B} \, \phi_0 \, \mathcal{B}^{-1} 
               + \mathcal{B}_x \, \mathcal{B}^{-1} \, ,  \label{ghier_BT1} \\
   \hchi_{n+1}(\mathcal{B})
 &=& - \phi_n' \, \mathcal{B} + \sum_{k=0}^n \hchi_k(\mathcal{B}) \, \phi_{n-k}   
     \qquad n=1,2,\ldots \; .  \label{ghier_BT2}
\ee

\noindent
{\bf Example 1.} Setting $\phi_n = -\hchi_n(\phi)$, $n=0,1, \ldots$, 
so that 
\be
         \mathcal{E}(\la) = I + \la \, \phi_{-[\la]} \, , 
\ee
the subsystem of (\ref{big_hier}) for $m=0$ reads 
\be
    \hchi_{n+1}(\phi) + \hchi_n(\phi_x + \phi^2) - \hchi_n(\phi) \, \phi = 0 
    \qquad n=0,1,\ldots \, ,
\ee
which in functional form, 
and after a Miwa shift, becomes the representation 
(\ref{oppBurgers}) of the `opposite' Burgers hierarchy. 
The remaining equations resulting from (\ref{big_hier}) are 
\bez
    \hchi_m \hchi_{n+1}(\phi) - \hchi_n \hchi_{m+1}(\phi) 
  = \sum_{k=1}^m \hchi_{m-k} \hchi_n(\phi) \, \hchi_k(\phi) 
    - \sum_{k=1}^n \hchi_{n-k} \hchi_m(\phi) \, \hchi_k(\phi) 
\eez
where $m,n=1,2,\ldots$. 
By use of the Hasse-Schmidt derivation property of the $\hchi_n$, 
this is the form (\ref{KP_hchi}) of the pKP hierarchy. 
But we already know that the latter is satisfied as a consequence of 
the Burgers hierarchy. 
Equations (\ref{gHC}) take the form 
\be
    \phi = - f_x \, f^{-1} \, , \qquad  
    \hchi_n(\phi) = \hchi_{n+1}(f) \, f^{-1} \qquad  n=1,2,\ldots \; .
\ee
This leads to the linear functional equation
\be
   f^{-1}_{[\la]} = f^{-1} + \la \, (f^{-1})_x  \, ,
\ee
and thus $\chi_n(f^{-1}) = 0$ for $n=2,3,\ldots$, which is equivalent 
to the following version of a linear heat hierarchy,
\be
   \pa_{t_n}(f^{-1}) = (-1)^{n+1} \pa_x^n(f^{-1}) \qquad n=2,3,\ldots \; . 
      \label{heat_hier2}
\ee 
As a consequence, if $f^{-1}$ solves the linear hierarchy (\ref{heat_hier2}), 
then $\phi = - f_x \, f^{-1}$ solves the Burgers hierarchy (\ref{oppBurgers}) 
and thus also the pKP hierarchy. 

Equations (\ref{ghier_BT1}) and (\ref{ghier_BT2}) are turned into
\be
   \phi' = \mathcal{B} \, \phi \, \mathcal{B}^{-1} - \mathcal{B}_x \, \mathcal{B}^{-1} 
       \, ,  \qquad
   (I + \la \, \phi') \, \mathcal{B}_{[\la]} = \mathcal{B} \, (I + \la \, \phi )
       \; . 
\ee
Using the first in the second equation to eliminate $\phi'$, yields 
an equation linear in $\mathcal{B}^{-1}$,
\be
   (\la^{-1} + \phi) ( \mathcal{B}^{-1}_{[\la]} - \mathcal{B}^{-1} )  
 = (\mathcal{B}^{-1})_x  \; . 
\ee
Comparison with the Burgers hierarchy system (\ref{oppBurgers}) shows that 
$\mathcal{B}^{-1} = \phi$ is a solution. More generally, 
$\mathcal{B}^{-1} = \alpha + \phi \, \beta$ with any constant $\alpha, \beta$ 
solves this equation.
\vskip.1cm

\noindent
{\bf Example 2.} Setting $\phi_n =0$ for $n>0$ and $\phi := \phi_0$, reduces the 
hierarchy (\ref{big_hier}) to the Burgers hierarchy of section~\ref{section:Burgers}, 
and the second of equations (\ref{gHC}) requires that $f$ has to solve the 
linear heat hierarchy. 
Relaxing the constraint to $\phi_n =0$ for $n>1$, thus leaving $\phi_0$ 
and $\phi_1$ as dependent variables, (\ref{big_hier}) results in 
\be
&&  (\hchi_{n+1}(\phi_0)-\hchi_n(\phi_0)\phi_0-\hchi_{n-1}(\phi_0)\phi_1) 
     \, \delta_{m,0}        \nonumber \\
&& + (\hchi_{n+1}(\phi_1)-\hchi_n(\phi_1)\phi_0-\hchi_{n-1}(\phi_1)\phi_1) 
     \, \delta_{m,1}        \nonumber \\
&=& (\hchi_{m+1}(\phi_0)-\hchi_m(\phi_0)\phi_0-\hchi_{m-1}(\phi_0)\phi_1) 
     \, \delta_{n,0}        \nonumber \\
&& + (\hchi_{m+1}(\phi_1)-\hchi_m(\phi_1)\phi_0-\hchi_{m-1}(\phi_1)\phi_1)
     \, \delta_{n,1} \; .    \label{restr_system}
\ee
It is sufficient to consider $m < n$. For $m=0,n=1$, this yields
\be
    \phi_{0,y}-\phi_{0,xx} - 2 \phi_{0,x} \phi_0 
 = 2 \phi_{1,x} + 2 [\phi_1,\phi_0] \; .
\ee
The remaining equations which result from (\ref{restr_system}) are ($m=0,n>1$)
\be
    \hchi_{n+1}(\phi_0) - \hchi_n(\phi_0) \, \phi_0 - \hchi_{n-1}(\phi_0) \, \phi_1 = 0 
    \qquad  n=2,3, \ldots \, ,
\ee
and ($m=1,n>1$)
\be
    \hchi_{n+1}(\phi_1) - \hchi_n(\phi_1) \, \phi_0-\hchi_{n-1}(\phi_1) \, \phi_1 = 0
    \qquad n=2,3,\ldots \; .
\ee
In the case under consideration, equations (\ref{gHC}) take the form 
\be
    \phi_0 = f_x \, f^{-1} \, , \qquad \phi_1 = - \hchi_2(f) \, f^{-1} 
         = {1 \over 2}(f_{y} - f_{xx}) \, f^{-1} \, ,
\ee
and 
\be
     \hchi_n(f) = 0 \, , \qquad  n = 3,4, \ldots \, , 
\ee
which is \emph{not} equivalent to the heat hierarchy since $\hchi_2(f)=0$ 
is missing.

\end{document}